\documentclass{elsart}
\usepackage{amssymb}

\usepackage{amsmath}
\usepackage[dvips]{graphicx}

\journal{\textbf{\ \ Continuum
Mechanics and Thermodynamics}, vol. \textbf{20}, n$^o$ 5.
\qquad\qquad\qquad\qquad\qquad\qquad\qquad\qquad\qquad\qquad }
\input{tcilatex}

\begin{document}

\begin{frontmatter}

\title{A new approach for  the limit to tree height using a liquid nanolayer model}

\author{Henri Gouin}
\ead{henri.gouin@univ-cezanne.fr}

\address {
 University of Aix-Marseille \&  C.N.R.S. U.M.R. 6181, \\ Case 322, Av. Escadrille
 Normandie-Niemen, 13397 Marseille Cedex 20 France}
\vskip 0.5cm
 \small{Published in \emph{Continuum Mechanics and Thermodynamics}.
 \\ Accepted after revision on 15 August 2008, in Vol.
 \textbf{20},
5
 (2008).\\
 On line first: 10 September 2008
 \\
 "The original publication is available at www.springerlink.com"
 \\  {\scriptsize DOI:   10.1007/s00161-008-0084-y }}

\begin{abstract}
Liquids in contact with solids are submitted to intermolecular
forces inferring   density gradients
   at the
walls. The van der Waals forces make liquid   heterogeneous, the stress tensor is not any more spherical as in homogeneous bulks and it is possible to obtain stable thin liquid films wetting
vertical walls up to altitudes that incompressible fluid models are
not forecasting. Application to micro tubes of xylem enables to
understand why the ascent of sap is possible for very high trees
like sequoias or giant eucalyptus.
\end{abstract}


\end{frontmatter}

\section{Introduction}
In \emph{Amazing numbers in biology}, Flindt reports  an eucalyptus of 128 meters and
    a giant sequoia of 135 meters \cite{Flindt}. However,
 biophysical
determination of maximum size to which  trees can grow is   not well
 understood  \cite{Koch}.
A main problem with the understanding of
tall trees is why the   sap is able to reach so high levels. \\
 Xylem tube  diameters   range between 50 and  400  $\mu m$; the crude sap  contains diluted salts
  but   its physical properties are roughly those of  water.
Consequently,
 hydrodynamic,  capillarity  and osmotic pressure
  create a sap ascent of   only  few tens of meters   \cite{Zimm}.
To explain the sap ascent phenomenon, Dixon and Joly proposed in
1894 a cohesion-tension model \cite{Dixon}, followed by a
quantitative attempt  by van der Honert in 1948 \cite{Honert}:
liquids may be subjected to tensions generating negative pressures
compensating gravity  effects. Nevertheless, thermodynamic states
are strongly metastable and can generate cavitation  causing
embolisms in  xylem tubes made of dead cells \cite{Tyree1}. \\ As
pointed out in Ref. \cite{Zimm2},  a turning-point in the pro and
con debate on the sap ascent was the  experiment of Preston in 1952
who demonstrated that tall trees survived overlapping double
saw-cuts made through the cross-sectional area of the trunk to sever
all xylem elements \cite{Preston}. This result, confirmed later by
several authors (e.g. Mackey \& Weatherley in 1973 \cite{Mackay};
Eisenhut, in 1988 \cite{Eisenhut}; Benkert \emph{et al} in 1991
\cite{Benkert}), was obviously not in agreement with the
cohesion-tension theory. Using a xylem pressure probe, Balling and
Zimmermann showed up that, in many circumstances, this apparatus
does not measure any water tension \cite{Balling}.  Since these
experiments, Zimmermann \emph{et al} questioned the cohesion-tension
theory in Ref. \cite{Zimmu}: xylem tension exceeding 0.6 Mpa was not
observed and in normal state most vessels were found to be embolized
at a level corresponding to sixty meter high
\cite{Tyree2}\footnote{It is interesting to remark that  xylem tube
diameters range between 50 and 400  $\mu m$ and the crude sap is a
liquid bulk with   a superficial tension $\sigma$ lower than the
superficial tension of pure water which is $72.5$ cgs at
20${{}^\circ}$ Celsius.  Let us consider a microscopic gaz-vapor
bubble inside the crude sap with a diameter $2\,R$ smaller than
xylem tube diameters. The difference between the gaz-vapor pressure
and the liquid sap pressure can be expressed by the Laplace formula:
$\displaystyle P_v-P_l = 2\,\sigma/R$. The vapor-gas pressure is
positive and omitted with respect to $\left|P_l\right|$;
consequently unstable bubbles appear when $\displaystyle
  R\geq
- 2\,\sigma/P_l$ . For a negative pressure  $P_l =-0.6$ MPa in the
sap as pointed out by experiments, we obtain  $R \geq 0.24\, \mu m$.
Due to the diameter range of   xylem tubes and the diameter range of
bubbles, the Laplace formula is valid at equilibrium for such bubble
sizes \cite{Isola}; dynamical bubbles appear spontaneously from
germs naturally included in the liquid  when the tubes are filled
with the crude sap and
 cavitation  makes the tubes embolized. }; consequently
trees growing taller than a few tens of meter range are not
foreseeable.
 Recently in their review article
Zimmermann \emph{et al} demonstrate that the present interpretation
of the pressure bomb data is based on a misconception and that
negative xylem pressure values of several megapascals do not exist
since xylem sap composition, the features of the xylem wall and the
hydraulic coupling of the xylem with the tissue prevent the
development of stable tensions larger than about 1 MPa. Moreover,
gas-vapor transportation in xylem tubes was found at the top of
height trees \cite{Zimm2}.

In this paper, our ambition is to present an understanding of the ascent of sap
in very high trees, different from the cohesion-tension theory:
 at
 a  higher level than a few
 tens of meters  - corresponding to the pulling of water by hydrodynamic, capillary and osmotic pressure -
 we assume  that  tubes may be
  embolized. In addition, we assume also that  a thin liquid  film  - with a thickness of a few
   nanometers  \cite{Derjaguin,Israel,Lifshitz} - wets   xylem
 walls up to the top of the tree. At this scale, long range molecular forces stratify  liquids \cite{Ono}  and
 the ratio between tube diameter  and
 sap film thickness allows us to consider  tube walls
 as \emph{plane surfaces}; consequently
 the problem of sap ascent in vertical
 tubes is similar to the rise of a liquid film damping a vertical plane
 wall. \newline The  sap motion in xylem tube can be   suitably explained by the transpiration through micropores located in tree leaves \cite{Zimm,Zimm2}:
  evaporation changes the liquid layer thickness implying  driving of sap as explained in Ref. \cite{gouin7}. \newline Consequently, this paper aims to prove, in non-evaporating case,
  the existence at equilibrium of thin films of water  wetting  vertical walls up to a same order of  altitude  than the height of very tall trees.

The recent development of experimental technics allows us to
 observe physical phenomena at length scales of a few nanometers. This
nanophysics reveals behaviors often surprising and basically different
from those that can be observed at a microscopic scale \cite{Bhushan}.
\newline
At the end of the nineteenth century, the fluid inhomogeneity in
liquid-vapor interfaces was taken into account by considering a
volume energy depending on space density derivative
\cite{vdW,Widom}. In the first part of the twentieth century, Rocard
obtained a thermodynamical justification of the model  by an
original step in kinetic  theory of gases \cite{Rocard}. This van
der Waals square-gradient functional  is unable to model repulsive
force contributions and misses the dominant damped oscillatory
packing structure of liquid  layers near a substrate wall
\cite{chernov1,chernov2}. Furthermore, the decay lengths are only
correct close to the liquid-vapor critical point where the damped
oscillatory structure is subdominant
 \cite{Evans1,Evans3}. Recently, in mean field theory, weighted density-functional has been used to explicitly demonstrate the dominance
 of this structural contribution in van der Waals thin films and to take into account long-wavelength capillary-wave fluctuations as in papers that
 renormalize the square-gradient functional to include capillary wave fluctuations \cite{Fisher,Henderson}. In contrast, fluctuations strongly damp
 oscillatory structure and it is mainly for this reason that van der Waals' original prediction of a \emph{hyperbolic tangent} is so close to simulations and experiments \cite{rowlinson}.\\
 To get an  analytic expression in  density-functional theory for a thin liquid film near a solid wall, we add a liquid density-functional
 at the solid    surface to the square-gradient functional  representing closely liquid-vapor interface free energy. This kind of functional
 is well-known in the literature,  as the  general background studied by Nakanishi and Fisher \cite{Fisher1}.
 It was used by Cahn in a phenomenological form, in a well-known paper studying wetting near a critical point \cite{Cahn0}.
 An asymptotic expression is obtained in \cite{gouin} with an approximation of hard sphere molecules and London potentials for
  liquid-liquid and solid-liquid interactions: by using London or Lennard-Jones potentials, we took into account the power-law
  behavior which is dominant in a thin liquid film in contact with a solid. In this paper, the effects of the vapor bulk bordering the liquid film
  are simply expressed  with an other density-functional of energy
  located on a mathematical surface as a dividing-like surface for  liquid-vapor interfaces of a few Angstr\"om thickness.
\\ With this functional,
we obtain the equations of equilibrium \cite{gouin4} and boundary
conditions \cite{Gouin1} for  a thin vertical liquid film damping a
vertical solid wall and we can compute the liquid layer thickness as
a function of the film level. Moreover, the normal stress vector
acting on the wall is constant through the liquid layer and
corresponds to the gas-vapor bulk pressure which is currently the
atmospheric
pressure; no negative pressure appears in the liquid layer. \\
As in \cite{seppecher}, several methods can be used to study the
stability of a thin liquid film in equilibrium. In our case, the
\emph{disjoining pressure} of a thin liquid layers is a well adapted
tool for very thin films. By using   Gibbs free energy per unit area
for the liquid layer as a function of the thickness, we are able to
obtain the minimal thickness for which a stable wetting film damps a
solid wall. The minimal thickness is associated with the
\emph{pancake layer} when the film is bordering  the dry solid wall
\cite{Derjaguin,de Gennes1,de Gennes2} and corresponds to a maximal
altitude. Numerical calculations associated with physical values for
water yield the maximal  film altitude for a silicon wall and a less
hydrophile material. In the two cases, we obtain an approximative
maximum level   corresponding to a good height order for the tallest
trees.

\section{Definition and well-known results of the disjoining pressure}
Without redoing or demonstrating the main results of Derjaguin \emph{et al} \cite{Derjaguin} related to
 thin liquid films and the well-known \emph{disjoining pressure}, we enumerate  the properties  we apply in the problem of  rise of  a liquid on a vertical wall.  \newline
In this paper, we consider  fluids and solids  \emph{at a given
temperature}  $\theta$. The film is thin enough such that the
gravity effect is neglected across the liquid layer. The hydrostatic
pressure in a thin liquid layer included between a solid wall and a
vapor bulk differs from the pressure in the contiguous liquid phase.
At equilibrium, the additional interlayer pressure is called the
\emph{disjoining pressure}.   Clearly, a disjoining pressure could
be measured by applying an external pressure to keep the complete
layer in equilibrium. The measure of a disjoining pressure is either
the additional pressure on the surface or the drop in the pressure
within the \emph{mother bulks} which produce the  layer. In both
cases, the forces arising during the thinning of a film of uniform
thickness $h$
 produce the disjoining pressure $\Pi(h)$ of the liquid layer with the surrounding phases; the disjoining pressure is equal to the difference between
  the pressure $P_{{v_b}}$ on the interfacial surface (which is the pressure of the \emph{vapor mother bulk} of density $\rho_{v_b}$) and the
  pressure $P_b$ in the \emph{liquid mother bulk} (density $\rho_b$) from which the liquid layer extends (this is the reason
  for which Derjaguin used the term \emph{mother bulk} \cite{Derjaguin}, page 32) :
\begin{equation}
\Pi(h) = P_{{v_b}}-P_b \,.\label{disjoiningpressure}
\end{equation}
The most classical apparatus to measure the disjoining pressure is
due to Sheludko \cite{Sheludko} and is described on Fig. (1).
\begin{figure}[h]
\begin{center}
\includegraphics[width=11cm]{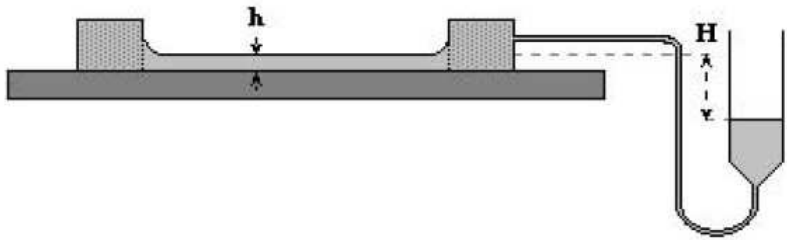}
\end{center}
\caption{\emph{Diagram of the technique for determining the
disjoining pressure isotherms of wetting films on a solid substrate:
a  circular wetting film is formed on a flat substrate to which a
microporous filter is clamped. A pipe connects the filter filled
with the liquid to a reservoir containing the liquid mother bulk
that can be moved by a micrometric device. As we will see in section
5, the thickness $h$ of the film depends on   $H$   in a convenient
domain of $H$ values where the wetting film is stable. The
disjoining pressure is equal to $\Pi = (\rho_{b}-\rho_{v_b})\,g\,H$,
where $g$ is the acceleration of gravity  (From Ref. \cite
{Derjaguin}, page 332).}}\label{fig1}
\end{figure}
Let us consider the Gibbs free energy  of the  liquid layer
(thermodynamic potential). As proved by Derjaguin \emph{et al} in
Ref. (\cite{Derjaguin}, {Chapter 2}),
 the   Gibbs free energy per unit area $G$ can be expressed as a function of
 $h$ :
\begin{equation}
\frac{\partial G(h)}{\partial h} = -\Pi(h). \label{Gibbs1}
\end{equation}
Eq. (\ref{Gibbs1}) can be integrated as :
\begin{equation}
G (h) = \int_{h}^{+\infty} \Pi(h)\,dh\label{Gibbs2}
\end{equation}
where $h=0$ is associated with the dry wall in contact with the
vapor bulk and $h=+\infty$ is associated with a wall in
 contact with   liquid bulk  when the value of $G$ is $0$.\\
 An important property related to the  problem of wetting
 is associated with the well-known spreading coefficient :
 \begin{equation*}
S = \gamma_{_{SV}} - \gamma_{_{SL}}-\gamma_{_{LV}},\label{wetting}
\end{equation*}
where  $\gamma_{_{SV}}, \gamma_{_{SL}}, \gamma_{_{LV}}$ are respectively the  solid-vapor, solid-liquid and liquid-vapor free energies per unit area of interfaces.
 The energy of the liquid layer per unit area can be written as :
 \begin{equation*}
E =  \gamma_{_{SL}}+\gamma_{_{LV}}+ G(h). \label{layer energy}
\end{equation*}
When $h=0$, we obtain the energy $\gamma_{_{SV}}$ of the dry solid
wall; when $h=+\infty$, we obtain $\gamma_{_{SL}} + \gamma_{_{LV}}$.
In complete wetting of a liquid on a solid wall, the spreading
coefficient is positive and the Gibbs free energy $G$ looks like in
Fig. 2.

\begin{figure}[h]
\begin{center}
\includegraphics[width=7.5cm]{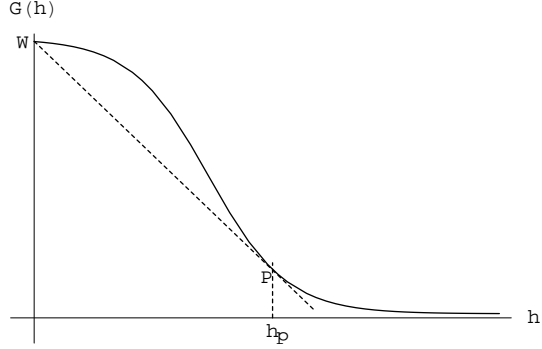}
\end{center}
\caption{\emph{ The construction of the tangent to  curve G(h) from point $W$ of coordinates $(0,G(0))$ involves point $P$; point
W  is associated with a high-energy surface  of the dry wall and point  P  is associated with   the pancake
  thickness $h_p$ where the film coexists with the dry wall; it is the smallest stable thickness  for the liquid layer.}}\label{fig2}
\end{figure}

The conditions of stability of a thin liquid layer essentially
depend on phases between which the film is sandwiched. In case of a
single film in equilibrium with the vapor and a solid substrate, the
stability condition is :
 \begin{equation}
 \frac{\partial \Pi (h)}{{\partial h}} < 0\quad\Leftrightarrow\quad
 \frac{\partial^2 G (h)}{{\partial h^2}} > 0 .\label{stability}
 \end{equation}
 The coexistence of two film segments with different thicknesses is a phenomenon which can be interpreted with
 the equality of chemical potential and superficial tension of the two films. A spectacular case
  corresponds to the coexistence of a liquid film of thickness $h_p$ and the dry solid wall
  associated with $h=0$.  The film is the so-called \emph{pancake  layer} corresponding to the
  condition :
  \begin{equation}
G(0) =  G(h_p)+ h_p \Pi(h_p). \label{pancake thickness}
\end{equation}
 Eq. (\ref{pancake thickness}) expresses that the value of the Legendre transformation of $G(h)$ at $h_p$ is equal to $G(0)$.
 Liquid films of thickness $h > h_p$ are stable and liquid films of thickness $h < h_p$ are metastable or unstable.

\section{The study of inhomogeneous fluids by using a square-gradient approximation and surface-density functionals at bordering walls}

The modern understanding of liquid-vapor interfaces begins with papers of van der Waals
and the square-gradient approximation for the free energy of inhomogeneous fluids. In current approaches,
it is possible to give formal exact expressions of the free energy in terms of pair-distribution function
and the direct correlation function \cite{Lutsko}. In practice, these complex expressions must be approximated to lead to a compromise between accuracy and simplicity.\\
 When we are confronted with such complications, the primitive mean-field models are generally
 inadequate and the obtained qualitative picture is no more sufficient. The main alternatives are
 density-functional theories which are a lot simpler than the Ornstein-Zernike equation
 in statistical mechanics since the local density is a functional at each point of the fluid \cite{Evans1,rowlinson}.
  We use this rough approximation enabling us to compute analytically the density profiles of simple fluids.
  Nevertheless, we take into account surface effects and repulsive forces by adding density-functionals at boundary surfaces.
The density-functional of the inhomogeneous fluid in a domain $O$ of
boundary $\partial O$  is taken in the form :
\begin{equation}
F = \int\int\int_O \varepsilon\ dv + \int\int_{\partial O} \varpi\
ds .\label{density functional}
\end{equation}
 The first integral is associated with a square-gradient approximation when we introduce a specific
free energy of the fluid at a given temperature $\theta$,
\begin{equation*}
\varepsilon =\varepsilon(\rho,\beta)
\end{equation*}
as a
function of density $\rho$ and  $\beta=
(\rm grad\, \rho)^2$. Specific free energy $\varepsilon$
characterizes together the fluid properties of \emph{compressibility} and
\emph{molecular capillarity} of liquid-vapor interfaces.
 In accordance  with gas kinetic
theory,
\begin{equation}
\rho \,\varepsilon =\rho \,\alpha (\rho)+\frac{\lambda
}{2}\,(\text{grad\ }\rho )^{2},  \label{internal energy}
\end{equation}
where term $ ({\lambda }/{2})\,(%
\mathrm{grad\ \rho )^{2}}$  is added  to the volume free energy
$\rho \,\alpha (\rho)$ of a compressible fluid and $\lambda =2\rho\,
\varepsilon _{\beta }^{\prime }(\rho,  \beta)$ is assumed to be constant
at a given temperature \cite{Rocard}. Specific free energy $\alpha $
enables to connect  continuously  liquid and vapor bulks and
 pressure $P(\rho)=\rho
^{2}\alpha _{\rho }^{\prime }(\rho )$ is similar to
van der Waals one.

Near a solid wall, London potentials of liquid-liquid and
liquid-solid interactions are :
\begin{equation*}
\left\{
\begin{array}{c}
\displaystyle\;\;\;\;\;\;\varphi
_{ll}=-\frac{c_{ll}}{r^{6}}\;,\text{ \ when\ }r>\sigma
_{l}\;\;\text{and }\;\ \varphi _{ll}=\infty \text{ \ when \ }r\leq
\sigma _{l}\, ,\  \\
\displaystyle\;\;\;\;\;\;\varphi
_{ls}=-\frac{c_{ls}}{r^{6}}\;,\text{ \ when\ }r>\delta
\;\;\text{and }\;\ \varphi _{ls}=\infty \text{ \ when \ }r\leq
\delta \;, \
\end{array}
\right.
\end{equation*}
where $c_{ll}$ and $c_{ls}$ are two positive constants associated
with Hamaker constants, $\sigma _{l}$ and $\sigma _{s}$ denote fluid
and solid  molecular diameters, $\delta =\frac{1}{2}($ $\sigma
_{l}+$ $\sigma _{s})$ is the minimal distance between centers of
fluid and solid molecules. Forces between liquid and solid  are
short range and can be described simply by adding a special energy
at the surface. This energy is the contribution to the solid/fluid
interfacial energy which comes from direct contact. This is not  the
entire interfacial energy: another contribution
 comes from the distortions in the density profile near the wall \cite{Cahn0,gouin,de Gennes1}.
 For a plane solid wall (at a molecular scale), this
  surface free energy is in the form :
\begin{equation}
\phi(\rho)=-\gamma _{1}\rho+\frac{1}{2}\,\gamma_{2}\,\rho^{2}. \label{surface energy}
\end{equation}
Here $\rho$ denotes the fluid density value  at surface $(S)$;
constants $\gamma _{1}$, $\gamma _{2}$ and $\lambda$ are positive
and given by the mean field approximation :
\begin{equation}
 \gamma
_{1}=\frac{\pi c_{ls}}{12\delta ^{2}m_{l}m_{s}}\;\rho _{sol},\quad
 \gamma _{2}=\frac{\pi c_{ll}}{12\delta^2 m_{l}^{2}},\quad  \lambda =   \frac{2\pi
c_{ll}}{3\sigma_l \,m_{l}^{2}},\label{coefficients}
\end{equation}
where $m_{l}$ and $m_{s}$ denote respectively masses of fluid and
solid molecules, $\rho _{sol}$ is the solid density \cite{gouin}.

We consider a  plane liquid layer contiguous to its vapor bulk and
in contact with a plane solid wall $(S)$; the z-axis is
perpendicular to the solid surface.
 The liquid film thickness is denoted by $h$; the
conditions in  the vapor bulk are  $\displaystyle {\rm grad}\, \rho
=0$ and
 $\Delta \rho = 0$ with $ \Delta$ denoting the Laplace operator.\\
 \emph{Far below from the critical point of the fluid},
 a way to compute the total free energy of the complete liquid-vapor
 layer is to add the surface energy of the solid wall $(S)$ at $z=0$, the energy of the liquid layer $(L)$
 located between $z=0$ and $z=h$, the energy of the sharp liquid-vapor
 interface of a few Angstr\"{o}m thickness assimilated to a surface $(\Sigma)$
  at $z=h$ and the energy of the vapor layer located between $z=h$ and
$z=+\infty$ \cite{Gavrilyuk}. The liquid at level $z=h$ is situated
at a distance order of two molecular diameters from the vapor bulk
and the vapor has a negligible density
 with respect to the liquid density \cite{Pismen}. \\
 In our model, the two last energies can be expressed with writing a unique
  energy $\psi$ per unit surface located on the  mathematical surface $(\Sigma)$ at  $z=h $ :
by a calculus like in Ref. \cite{gouin}, we can write $\psi$ in the same
form than Rel. (\ref{surface energy}) and also expressed as
in Ref. \cite{de Gennes1} in the form $\psi(\rho) = -\gamma
_{5}\rho+\frac{1}{2}\,\gamma_{4}\,\rho^{2}$; but with a \emph{wall}
corresponding to a \emph{negligible density}, $\gamma _{5}\simeq 0$,
the surface free energy $\psi$  is reduced to :
\begin{equation}
\psi (\rho )=\frac{\gamma _{4}}{2}\ \rho ^{2}, \label{cl2}
\end{equation}

where $\rho$ is the liquid density at level $z=h$ and  $ \gamma_4$
is associated with a distance $d$ of the  order of the fluid molecular
diameter (then $d \simeq  \delta$ and  $\gamma_4\simeq \gamma_2 $).
 Consequently, due to the small
vapor density,  the surface free energy $\psi$  is the same than the
surface free energy of a liquid in contact with a vacuum.\\
Complementary to this argumentation,  we will see, in section 4,
that the boundary condition at surface ($\Sigma$) associated with
surface energy (\ref{cl2}) yields a density value corresponding to
an intermediate density between liquid and vapor and which can be
considered as a density value of a dividing-like surface separating
liquid and vapor inside the
liquid-vapor interface.\\
Density-functional (\ref{density functional}) of the liquid-vapor
layer gets the final form :
\begin{equation}
F = \int\int\int_{(L)} \varepsilon\ dv + \int\int_{(S)} \phi\ ds +
\int\int_{(\Sigma)} \psi\ ds. \label{density functional2}
\end{equation}

\section{Equation of equilibrium and boundary conditions of a thin liquid layer contiguous to its vapor bulk and in contact with a vertical plane  solid wall}

In case of equilibrium,   functional  (\ref{density functional2}) is
stationary and  yields the \emph{equation of equilibrium} and the
\emph{boundary conditions} \cite{Gouin1,seppecher,Pismen}.
\subsection{Equation of equilibrium}
 The equation of equilibrium  is :
\begin{equation*}
 \text{div }\mathbf{\sigma } -\rho \text{
grad }\Omega   = 0 ,\label{motion0}
\end{equation*}
where $\Omega $ is the body force potential and $\mathbf{\sigma }$  the
stress tensor generalization  \cite{trusk,gouin4},
\begin{equation*}
\mathbf{\ \sigma =}-p\,\mathbf{1}-\lambda \;\text{grad\ }\rho \
\otimes \ \text{grad }\rho ,  \label{contrainte}
\end{equation*}
with   $p=\rho ^{2}\varepsilon _{\rho }^{\prime }-\rho \text{ div\textrm{\ }}%
(\lambda \text{ grad }\rho )$.\\
Let us consider   an isothermal  vertical film
of liquid bounded respectively by a flat solid wall and a vapor
bulk; then
\begin{equation}
\text{div }\mathbf{\sigma} + \rho\, g\, \mathbf{i}=0\,
\label{equilibrium1a}
\end{equation}
in orthogonal system, where $\mathbf{i}$ is the downward direction
of coordinate
 $x$ (the gravity potential   is $\Omega = -
g\,x$).\\ The coordinate  $z$ being external and  normal to the flat vertical solid
wall, spatial density derivatives  \emph{are negligible} in
directions other than direction of $z$  corresponding to a very
strong gradient of density of the liquid normally to the layer  and a
weak inhomogeneity along the film. In the complete liquid-vapor
layer (we call \emph{interlayer}),
\begin{equation*}
\mathbf{\sigma}=
\left[
\begin{array}{ccc}
a_1, & 0, & 0 \\
0, & a_2, & 0 \\
0, & 0, & a_3
\end{array}
\right],\quad {\rm with} \quad \left\{
\begin{array}{ccc}
a_1=a_2 = \displaystyle -P+\frac{\lambda }{2}\left(\frac{d\rho }{dz}\right)^2+\lambda\, \rho\,
\frac{d^2\rho}{dz^2} \\
a_3= \displaystyle -P-\frac{\lambda }{2}\left(\frac{d\rho }{dz}\right)^2+\lambda\, \rho\,
\frac{d^2\rho}{dz^2}
\end{array}\right.
\end{equation*}
and Eq. (\ref{equilibrium1a}) yields a constant value  at level $x$
for the eigenvalue $a_3$,
\begin{equation*}
P+\frac{\lambda }{2}\left(\frac{d\rho }{dz}\right)^2-\lambda\, \rho\,
\frac{d^2\rho}{dz^2}=P_{v_{b_x}},   \label{equilibrium1b}
\end{equation*}
where $P_{v_{b_x}}$ denotes  the pressure $P(\rho_{v_{b_x}})$ in the
vapor bulk of density $\rho_{v_{b_x}}$ bounding the liquid layer at
level $x$. In the interlayer, eigenvalues $a_1, a_2$ are not
constant but depend on the distance $z$ to the solid wall. In all
the fluid, Eq. (\ref{equilibrium1a}) can also be written
\cite{gouin4} :
\begin{equation}
 \text{grad }\left( \,\mu   -\lambda\, \Delta
\rho - g\,x\, \right) =0 \,, \label{equilibrium2a}
\end{equation}
where $\mu $ is the chemical potential (at a temperature $\theta $)
defined to an unknown additive constant.  We note that Eqs.
(\ref{equilibrium1a}-\ref{equilibrium2a}) are independent of surface
energies   (\ref{surface energy}) and (\ref{cl2}). \\The chemical
potential is a function of $P$ (and $\theta$); due to the equation
of state for pressure $P$, the chemical potential can be also
expressed as a function of $\rho$ (and $\theta$).  We choose as
\emph{reference chemical potential} $\mu _{o}=\mu _{o}(\rho)$ null
for bulks of densities $\rho _{l}$ and $\rho _{v}$ of phase
equilibrium. Due to Maxwell
 rule,
the volume free energy associated with $\mu _{o}$ is
$g_{o}(\rho)-P_{o}$  where   $ P_{o}=P(\rho _{l})=$ $P(\rho _{v})$
is the bulk pressure  and $g_{o}(\rho)=\displaystyle \int_{\rho
_{v}}^{\rho }\mu _{o}(\rho)\,d\rho\ $ is   null for the liquid and
vapor  bulks of phase equilibrium. The pressure $P$ is :
\begin{equation}
P(\rho)=\rho \, \mu _{o}(\rho)-g_{o}(\rho)\ +P_{o} .\label{therm.pressure}
\end{equation}
Thanks to Eq. (\ref{equilibrium2a}), we obtain in all the fluid
\emph{and not only in the interlayer} :
\begin{equation*}
\mu _{o}(\rho)-\lambda \Delta \rho -   g\ x =\mu
_{{o}}(\rho _{b}) ,\label{equilibrium2b}
\end{equation*}
where $\mu _{{o}}(\rho _{b})$ is the chemical potential
value of  a liquid mother bulk  of density  $\rho _{b}$
 such that $\mu _{{o}}(\rho _{b})= \mu
_{{o}}(\rho_{v_{b}})$, where $\rho_{v_{b}}$ is the density of the
vapor mother bulk bounding the layer \emph{at level} $x =0$. This
property is due to Eq. (\ref{equilibrium2a}) which is valid not only
in the liquid   but also in all the fluid   independently of the
surface energies in the density-functional (\ref{density
functional2}). Equation (\ref{equilibrium2a}) is also valid in the
sharp liquid-vapor interface.
\\  We
must emphasize that  $P(\rho _{b})$ and $P(\rho_{v_{b}})$ are unequal
as for drop or bubble  bulk pressures. Likewise, we define  a liquid
mother bulk  of density $\rho_{{b_x}}$
 at level  $x$ such that $\mu _{{o}}(\rho _{b_x})= \mu
_{{o}}(\rho_{v_{b_x}})$ with $P(\rho _{b_x}) \neq
 P(\rho_{v_{b_x}})$;     $\rho _{b_x}$ is
  not a  fluid density in the liquid layer but density in the liquid bulk
  from which the interlayer can extend. Then,
 \begin{equation}
\lambda\, \Delta \rho  =\mu_o(\rho)-\mu_o(\rho _{b_x}) \ \ {\rm with}\ \ \mu_o(\rho _{b_x}) = \mu_o(\rho _{b}) + g\,x   \label{equilibrium2c}
\end{equation}
and in the interlayer
 \begin{equation}
\lambda\,\frac{d^2\rho}{dz^2} = \mu_{b_x}(\rho), \quad {\rm with}\quad
\mu_{b_x}(\rho) = \mu_o(\rho)-\mu_o(\rho _{b_x})
\label{equilibrium2d}
\end{equation}
\subsection{Boundary conditions}
The condition  at the solid wall $(S)$   associated with  the free
surface energy  (\ref{surface energy}) yields \cite{Gouin1}:
\begin{equation}
\lambda \left(\frac{d\rho }{dn}\right)_{|_S}+\phi ^{\prime }(\rho)_{|_S}\ =0,
\label{cl1}
\end{equation}
where $n$ is the external normal direction to the fluid.  Eq.
(\ref{cl1}) yields :
\begin{equation*}
\lambda \left(\frac{d\rho }{dz}\right)_{|_{z=0}}=-\gamma _{1}+\gamma
_{2\ }\rho_{|_{z=0}} . \label{BC1}
\end{equation*}
The condition at the liquid-vapor interface  $(\Sigma)$  associated with
the free surface energy (\ref{cl2}) yields :
\begin{equation}
\lambda \left(\frac{d\rho }{dz}\right)_{|_{z=h}}=-\gamma _{4}\
\rho_{|_{z=h}}\,.  \label{BC2}
\end{equation}
As we will see in Sects. 5 and 6, Rel. (\ref{BC2}) takes into
account the density at $z=h$ which is smaller, but of the same order,
than liquid density. Due to the numerical values of $\lambda$ and
$\gamma_4$ in Sect. 6, the density derivative $\,\displaystyle
\frac{d\rho }{dz}\,$ is large with respect to the variations of the
density in the interlayer and corresponds to the drop of density in
the liquid-vapor interface in  continuous model. Consequently, Rel.
(\ref{BC2}) defines the film thickness by introducing a reference
point inside the liquid-vapor interface bordering the liquid layer
with a convenient density at surface $z=h$ considered as a kind of
dividing-like surface in a continuous model for the liquid-vapor
interface (\cite{rowlinson}, Chapter 3).

\section{The disjoining pressure for vertical liquid films}

Eq. (\ref{disjoiningpressure}) can be extended with the disjoining
pressure at level $x$ \cite{Derjaguin} :
\begin{equation*}
\Pi =P_{v_{b_x}}-P_{b_x}\,,  \label{disjoining pressure}
\end{equation*}
where $P_{b_x}$ and $P_{v_{b_x}}$ are the pressures in liquid and
vapor mother bulks  corresponding to   level $x$. At a given
temperature $\theta$,   $\Pi$ is a function of $\rho_{b_x}$ or
equivalently a function of $x$. Let us denote by
\begin{equation}
g_{b_x}(\rho) =
g_o(\rho)-g_o(\rho_{b_x})-\mu_{o}(\rho_{b_x})(\rho-\rho_{b_x}),\label{g}
\end{equation}
the primitive of  $\mu _{b_x}(\rho)$ null for $\rho _{b_x}$.
Consequently, from Eq. (\ref{therm.pressure}),
\begin{equation}
\Pi (\rho _{b_x}) = -g_{b_x}(\rho_{v_{b_x}}) , \label{disjoining}
\end{equation}
and an integration of Eq. (\ref{equilibrium2d}) yields :
\begin{equation}
\frac{\lambda }{2}\,\left(\frac{d\rho }{dz}\right)^2=g_{b_x}(\rho)+\Pi (\rho _{b_x}).  \label{equilibrium2e}
\end{equation}
The reference chemical potential linearized near $\rho_l$ is $\ \mu
_{o}(\rho)=\displaystyle  \frac{c_{l}^{2}}{\rho _{l}}(\rho
-\rho_{l})\ $ where $c_l$ is the isothermal sound velocity in liquid
bulk  $\rho_l$ at temperature $\theta$ \cite{espanet}. In the liquid
part of the liquid-vapor film,
 Eq. (\ref{equilibrium2d}) of density profile yields :
\begin{equation}
\lambda \frac{d^{2}\rho }{dz^{2}} =\frac{c_{l}^{2}}{\rho _{l}}(\rho
-\rho _{b})-g\,x \equiv \frac{c_{l}^{2}}{\rho _{l}}(\rho
-\rho _{b_x})\quad  {\rm with}\ \ \rho _{{b_x}} = \rho _{{b}}+  \frac{\rho _{l}}{c_l^2}\, g\, x .\label{liquidensity}
\end{equation}
The reference chemical potential linearized near $\rho_v$ is $\ \mu
_{o}(\rho)=\displaystyle  \frac{c_{v}^{2}}{\rho _{v}}(\rho
-\rho_{v})\ $ where $c_v$ is the isothermal sound velocity in vapor
bulk $\rho_v$ at temperature $\theta$ \cite{espanet}. In the vapor
part    of the liquid-vapor film,
\begin{equation*}
\lambda \frac{d^{2}\rho }{dz^{2}}=\frac{c_{v}^{2}}{\rho _{v}}(\rho
-\rho _{v_{b}})-g\,x\equiv \frac{c_{v}^{2}}{\rho _{v}}(\rho -\rho
_{v_{b_x}})\quad {\rm with}\quad  \rho _{v_{b_x}} = \rho _{v_{b}}+
\frac{\rho_v}{c_v^2}\,g\,x .
\end{equation*}
Due to Eq. (\ref{equilibrium2c}), $\mu _{o}(\rho)$ has the  same
value for $\rho _{v_{b_x}}$ and $\rho _{{b_x}}$, then
\begin{equation*}
\frac{c_{l}^{2}}{\rho _{l}}(\rho _{b_x}-\rho _{l}) =\mu _{o}(\rho
_{b_x})=\mu _{o}(\rho _{v_{b_x}})
=\frac{c_{v}^{2}}{\rho _{v}}(\rho _{v_{b_x}}-\rho _{v}), \ \  \rm and
\label{densities1}
\end{equation*}
\begin{equation*}
\rho _{v_{b_x}}=\rho _{v}\left(
1+\frac{c_{l}^{2}}{c_{v}^{2}}\frac{(\rho _{b_x}-\rho _{l})}{\rho
_{l}}\right) .  \label{densities2}
\end{equation*}

In  the liquid and vapor parts of the interlayer we have,
\begin{equation*}
g_{o}(\rho)=\frac{c_{l}^{2}}{2\rho _{l}}(\rho -\rho
_{l})^{2}\ \ \ {\rm ( liquid)}\quad {\rm and} \quad g_{o}(\rho)=\frac{c_{v}^{2}}{2\rho _{v}}(\rho -\rho
_{v})^{2}\ \ \ {\rm ( vapor)}.
\end{equation*}
From Eqs (\ref{g})-(\ref{disjoining})  we deduce immediately the disjoining pressure at level $x$
\begin{equation}
\Pi (\rho _{b_x})=\frac{c_{l}^{2}}{ 2\rho _{l}}(\rho _{l}-\rho _{b_x})%
\left[ \rho _{l}+\rho _{b_x}-\rho _{v}\left( 2+\frac{c_{l}^{2}}{c_{v}^{2}}%
\frac{(\rho _{b_x}-\rho _{l})}{\rho _{l}}\right) \right] .
\label{disjoining pressure2}
\end{equation}
Due to\ $\ \displaystyle\rho _{v}\left( 2+\frac{%
c_{l}^{2}}{c_{v}^{2}}\frac{(\rho _{b_x}-\rho _{l})}{\rho
_{l}}\right) \ll \rho _{l}+\rho _{b_x}$, we get
$$\displaystyle \Pi
(\rho _{b_x})\approx\frac{c_{l}^{2}}{2\rho _{l}}(\rho
_{l}^{2}-\rho _{b_x}^{2}) . $$
 At level $x=0$,   the liquid mother
bulk density is closely equal to  $\rho_l$  (density of liquid in
phase equilibrium) and because of Eq. (\ref{liquidensity}), $\Pi$
can be considered as a function of $x$ :
\begin{equation}
\Pi (x)=-\rho_l\,g\,x\left(1+\frac{g\,x}{2\,c_l^2}\right).
\label{disjoining pressuregravity}
\end{equation}
Now, we consider a film   of thickness  $h_x$ at level $x$; the
density profile in the liquid part of the liquid-vapor film is
solution of  :
\begin{equation}
\left\{
\begin{array}{c}
\displaystyle\lambda \frac{d^{2}\rho }{dz^{2}}=\frac{c_{l}^{2}}{\rho _{l}}
(\rho -\rho _{b_x}),  \\
\quad  {\rm with}\quad \displaystyle\lambda \frac{d\rho
}{dz}_{\left| _{z=0}\right. }=-\gamma _{1}+\gamma _{2\ }\rho
_{\left| _{z=0}\right. }\quad {\rm and}\quad
\displaystyle\lambda \frac{d\rho }{dz}_{\left| _{z=h_x}\right.
}=-\gamma _{4}\ \rho _{\left| _{z=h_x}\right. }.
\end{array}
\right.  \label{systeme1}
\end{equation}
Quantities $\tau $ and $%
d $ are defined such that : \begin{equation}\tau \equiv
\frac{1}{d}=\frac{c_{l}}{\sqrt{\lambda \rho _{l}}}\ , \label{tau}
\end{equation} where $d $ is a  reference length   and   we
introduce   coefficient $ \gamma _{3}\equiv\lambda \tau$. The solution
of system (\ref{systeme1}) is :
\begin{equation}
\rho =\rho _{b_x}+\rho _{1_x}\,e^{-\tau z}+\rho _{2_x}\,e^{\tau z},
\label{profil}
\end{equation}
where the boundary conditions at $z=0$ and $h_x$ yield the values of $\rho _{1_x}$ and
$\rho _{2_x}$:
\begin{equation*}
\left\{
\begin{array}{c}
(\gamma _{2}+\gamma _{3})\rho _{1_x}+(\gamma _{2}-\gamma _{3})\rho
_{2_x}=\gamma
_{1}-\gamma _{2}\rho _{b_x}, \\
\ \ \ \quad \quad -e^{-h_x\tau }(\gamma _{3}-\gamma _{4})\rho
_{1_x}+e^{h_x\tau }(\gamma _{3}+\gamma _{4})\rho _{2_x}=-\gamma _{4}\rho
_{b_x}.
\end{array}
\right. \
\end{equation*}
The liquid density profile is a consequence of Eq. ({\ref{profil})
when   $z$ $\in \left[ 0,h_x\right]$.\newline By taking Eq.
({\ref{profil}) into account  in Eq. (\ref {equilibrium2e}) and
$g_{b_x}(\rho)$ in  linearized form in the liquid part of the
interlayer, we get immediately
\begin{equation*}
\Pi (\rho _{b_x})=-\frac{2\,c_{l}^{2}}{\rho _{l}}\,\rho _{1_x}\,
\rho _{2_x},
\end{equation*}
 and   consequently,
\begin{eqnarray}
\Pi (\rho _{b_x}) &=&\frac{2\,c_{l}^{2}}{\rho _{l}}\left[ (\gamma
_{1}-\gamma _{2}\rho _{b_x})(\gamma _{3}+\gamma _{4})e^{h_x\tau
}+(\gamma
_{2}-\gamma _{3})\gamma _{4}\rho _{b_x}\right] \times  \notag \\
&&\frac{\left[ (\gamma _{2}+\gamma _{3})\gamma _{4}\rho
_{b_x}-(\gamma
_{1}-\gamma _{2}\rho _{b_x})(\gamma _{3}-\gamma _{4})e^{-h_x\tau }\right] }{%
\left[ (\gamma _{2}+\gamma _{3})(\gamma _{3}+\gamma _{4})e^{h_x\tau
}+(\gamma _{3}-\gamma _{4})(\gamma _{2}-\gamma _{3})e^{-h_x\tau
}\right] ^{2}} .\label{Derjaguine}
\end{eqnarray}
By identification of expressions  (\ref{disjoining pressure2}) and
(\ref {Derjaguine}), we get a relation between $h_x$ and $\rho
_{b_x}$ and consequently  a relation between  disjoining pressure
$\Pi (\rho _{b_x})$ and  thickness  $h_x$ of the liquid film. For
the sake of simplicity, we denote finally the disjoining pressure by
 $\Pi (h_x)$
which is a function of $h_x$ at temperature $\theta$.
\newline
Due to the fact that $\rho _{b_x}\simeq \rho _{b}\simeq \rho _{l}$
\cite{Derjaguin}, the disjoining pressure reduces to the simplified
expression :
\begin{eqnarray}
\Pi (h_x) &=&\frac{2\,c_{l}^{2}}{\rho _{l}}\left[ (\gamma
_{1}-\gamma _{2}\rho _{l})(\gamma _{3}+\gamma _{4})e^{h_x\tau
}+(\gamma _{2}-\gamma
_{3})\gamma _{4}\rho _{l}\right] \times  \notag \\
&&\frac{\left[ (\gamma _{2}+\gamma _{3})\gamma _{4}\rho
_{l}-(\gamma
_{1}-\gamma _{2}\rho _{l})(\gamma _{3}-\gamma _{4})e^{-h_x\tau }\right] }{%
\left[ (\gamma _{2}+\gamma _{3})(\gamma _{3}+\gamma _{4})e^{h_x\tau
}+(\gamma_{3}-\gamma _{4})(\gamma _{2}-\gamma _{3})e^{-h_x\tau
}\right] ^{2}} .\label{Derjaguine bis}
\end{eqnarray}
\emph{Let us notice  an important property}   of a mixture of a van
der Waals fluid and a perfect gas where the total pressure is the
sum of the partial pressures of components \cite{espanet}: at
equilibrium, the partial pressure of the perfect gas is constant
through the liquid-vapor-gas layer -where  the perfect gas is
dissolved in the liquid. The  disjoining pressure of the mixture is
the same than for a single van der Waals fluid and calculations and
results are identical to those previously obtained.

\section{Numerical calculations for water wetting a vertical plane wall}

Our aim is not to propose an exhaustive study of  the disjoining
pressure of water  for all  physicochemical conditions associated
with different walls  but to point out examples such that previous
results provide   new values of maximum height  for a vertical water
film damping a plane wall.\newline Calculations are made with
$Mathematica^{TM}$. The disjoining pressure $\Pi$ and the Gibbs free
energy $G$ are presented as functions of $h_x$. The $h_x$ values
must be greater than the molecular radius of water corresponding to
the smallest thickness of the liquid layer.\\
The graphs of $\Pi(h_x)$ are directly issued from Rel.
(\ref{Derjaguine bis}) with different physical values obtained in
the literature. The graphs of $G(h_x)$ are deduced from Rel.
(\ref{Gibbs2}). As a function of $h_x$, function $\Pi(h_x)$ is not
an analytically integrable expression; consequently G-graphs are
computed by $Mathematica^{TM}$ but with help of a numerical process.

 For a few nanometer
range, the film thickness is not  exactly $h_x$; at this range we
must add to $h_x$ the liquid part of the liquid-vapor interface
bordering the liquid layer (the thickness of which is neglected for
films of several nanometers). We can estimate this  part thickness
at $2\,\sigma_l$ (half of the thickness   of water liquid-vapor
interface at $ 20 {{}^\circ}  $ Celsius \cite{Rocard,rowlinson}).
The film thickness is   $e_x \approx h_x+ 2\, \sigma_l$. The
previous results  in sections 3-5 remain unchanged by   using
$h_x$ in place of the liquid thickness $h$.

 When $h_x = 0$ (corresponding to the dry
wall), the value of $G$ is the spreading coefficient $S$ (see Fig.
2). We must emphasize that point  $P$ associated with the pancake
layer is observed, on the numerical curves,  to be closely an
inflexion point of graph $\Pi(h)$ corresponding to the strongest
stability of  films (maximum of $\partial ^2 G/\partial h^2$)
\cite{de Gennes1,de Gennes2}. To obtain the pancake   thickness
corresponding to the smallest film thickness, we draw the graphs of
$\Pi(h_x)$ and $G(h_x)$, when $h_x \in
[\frac{1}{2}\,\sigma_l,\ell]$, where $\ell$ is a distance of few
tens of Amgstr\" om.

At  $\theta= 20 {{}^\circ}$ Celsius, we consider successively water
wetting a  wall in silicon as a reference of material  and water
damping a less wetting  wall.

 In  \textbf{c.g.s.  units}

  The
experimental estimates  of coefficients are obtained in Refs.
\cite{Israel} and \cite{Handbook} :
\newline
$\rho_l =0.998$,\newline $c_l = 1.478\times 10^{5}$,\newline
$c_{ll}=1.4\times 10^{-58}$,\newline $\sigma _{l}=2.8\times
10^{-8}\text{\ (2.8 Angstr\"{o}m or 0.28 nanometer), }$\newline
$m_{l}=2.99\times 10^{-23}$  .

 From Rel. (\ref{coefficients}), we deduce
\newline $\lambda =1.17\times 10^{-5}$,
\newline    $\gamma _2=\gamma _4 = 54.2$ .\newline
 From $\gamma _3 =\lambda\tau$ and Rel. (\ref{tau}), we get \newline $\gamma _3
= 506$,
\newline $d = 2.31 \times 10^{-8}$.

\emph{We consider two cases} :

\hskip 0.5cm \emph{a)} For silicon (as a reference of wall damped by
water), physical characteristics are,\newline $\sigma _{s}=2.7\times
10^{-8}$,\newline $m_{s}=4.65\times 10^{-23}$, \newline
$\rho_{sol}=2.33$ .  \newline No information is available for
water-silicon interactions; if  we assume that
$c_{ll}=c_{ls}=1.4\times 10^{-58}$, we deduce \newline    $\gamma
_1=81.2$ .

\hskip 0.5cm \emph{b)}  We consider  a material    such that
$\gamma_1 = 75$  (the material is less damped by liquid water). The
other values of the coefficients of the material are assumed to be the
same than in case \emph{a)}.  We will see that these values are well
adapted to our problem.

Corresponding to Rel. (\ref{stability}), the graphs of $\Pi(h_x)$ in
cases \emph{a)} and \emph{b)} are easy to plot following the liquid
layer thickness for stable and unstable domains. The \emph{G}-graphs
are deduced by numerical integration following the bound $h_x$;  the
limit $+\infty$ is replaced by ten thousand molecular diameters of
water molecules. Due to $h_x>\frac{1}{2} \,\sigma_l$, it is not
possible to obtain numerically the limit point $W$ corresponding to
the dry wall (Fig. 2). This point is graphically obtained by an
interpolation associated with the concave part of the
\emph{G}-curve. Point $P$ follows from the drawing of the tangent
line from $W$ to the \emph{G}-curve.

It is important to point out that  reference length $d$ is of the
same order than $\sigma_l,\, \sigma_s$ and $\delta$ and seems a good
length order for very thin films.\\
In Fig. 3, we present    disjoining pressure graphs in the two
cases. Real parts of  disjoining pressure graphs corresponding to
$\partial\Pi/\partial h_x <0$   are plain lines of the curves and
are associated with thickness  liquid layers of several molecules.
Dashed lines of the curves have no real existence.

From reporting the pancake thickness of the $h_x$ axis for the
$\Pi$-curve, we deduce its disjoining pressure corresponding value;
the maximum of altitude of topmost trees is calculated with Eq.
(\ref{disjoining pressuregravity}).

\begin{figure}[h]
\begin{center}
\includegraphics[width=11cm]{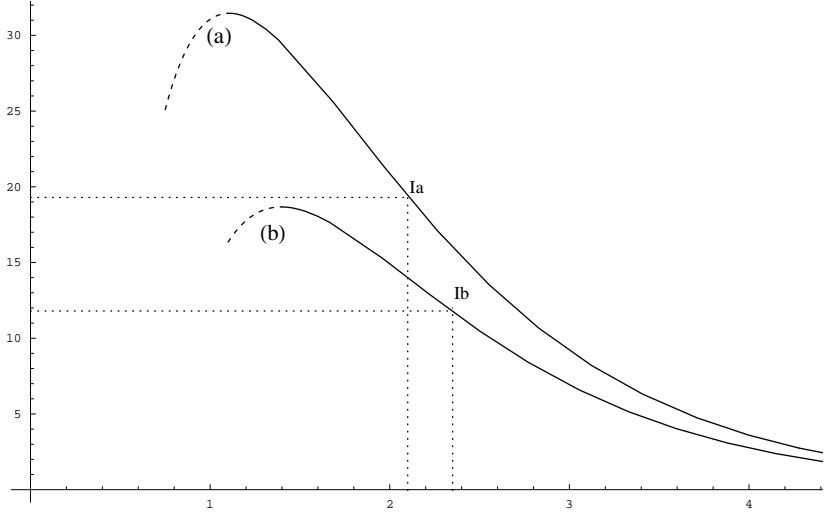}
\end{center}
\caption{\emph{Graphs represent the disjoining pressure as a
function of the  liquid layer thickness $h_x$  for liquid water at
$20{{}^{\circ }}$ C\ \ in contact with different walls (the total
thickness of the film is $e_x \approx h_x+ 2 \,\sigma_l$). The unit
of $x-$axis is $d=2.31\times 10^{-8}$ \texttt{cm}; the unit of
$y-$axis is one atmosphere. Curve (a) corresponds to water wetting a
silicon wall ($\gamma_1 = 81.2$ \texttt{cgs}); curve (b) corresponds
to water wetting a less damped wall ($\gamma_1 = 75$ \texttt{cgs}).
Points Ia and Ib are the inflexion points of curves (a) and (b)
corresponding roughly to points P in Fig. 4.}} \label{fig1}
\end{figure}
In Fig. 4, we present  graphs of the Gibbs free energy $G$ as a function of $h_x$.
The limit  of the film thickness is associated to the \emph{pancake}  thickness
 $e_p\approx h_p + 2\,\sigma_l$ when the liquid film coexists with the dry wall.
 The spreading coefficient values   are   associated with point $P$:
\begin{figure}[h]
\includegraphics[width=15cm]{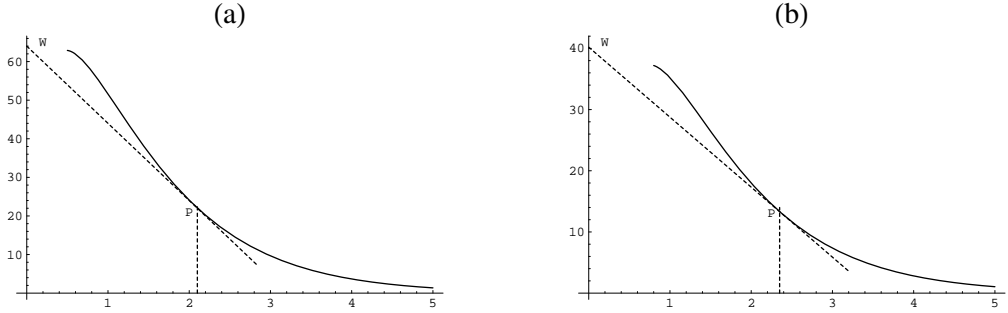}
\caption{\emph{Graphs represent the film Gibbs free energy  per unit
area as a function of the  liquid layer thickness for liquid water
at $20{{}^{\circ }}$ C in contact with two different walls. The unit
of $x-$axis is $d=2.31\times 10^{-8}$ \texttt{cm}; the unit of
$y-$axis is one \texttt{cgs} unit of superficial tension. Graph (a)
corresponds to water wetting a   silicon wall ($\gamma_1 = 81.2$
\texttt{cgs}); graph (b) corresponds to water  wetting a less damped
wall ($\gamma_1 = 75$ \texttt{cgs}). Point  W  is associated with
the surface energy of the dry wall and point  P  is associated with
$h_p$ corresponding to  the pancake layer where the film coexists
with the dry wall; the smallest film thickness possible is $e_p
\approx h_p+ 2 \sigma_l$.}} \label{fig2}
\end{figure}
at point $P$,  which is numerically close to the inflexion point of $\Pi(h_x)$,  the tangent  goes to point $W$ of the $y$-axis (where $G(0)=S$). In the two cases, the total pancake thickness $e_p=h_p+2\,\sigma_l$ is of one nanometer order corresponding to a good thickness value for a high-energy surface \cite{de Gennes2}.\\
From the graphs, we deduce  $S \approx 64$ \texttt{cgs} in case
\emph{a)} and   $S \approx 40$ \texttt{cgs }in case  \emph{b)}
corresponding to a less energetic wall.
 However, crude sap is not pure water.  Its liquid-vapor surface tension has a  lower value than surface tension of pure water (72 \texttt{cgs} at $20{{}^{\circ }}$C) and it is possible to obtain the same spreading coefficients with less energetic surfaces.
\\
When $|x|$ is of some hundred meters,  Eq. (\ref{disjoining
pressuregravity}) yields :
\begin{equation*}
\Pi(x)\simeq \rho_l\,g\,x .
\end{equation*}
The maximum of altitude $|x_{_M}|$ corresponds to the pancake layer.
We add  20 meters to this altitude, corresponding to the ascent of
sap due to hydrodynamic,  capillarity  and osmotic pressure. In case
 \emph{b)}, the material has a lower surface energy than silicon and we
obtain a film  height of    $140$ meters.

\section{Conclusion}

In this article, we considered a very thin liquid film damping
 and rising along a vertical plane wall as a model of the ascent of sap in xylem tubes.\newline
 At its bottom, the
  film is bordered on a liquid meniscus corresponding to  a xylem tube filled with sap up to an altitude of a few ten meters.
  Above this altitude, the xylem tube is embolized such as  the  liquid water thin film wets the wall of the tube.\newline
   The   study is static. The motor of the sap motion is induced by
   the transpiration across micropores located in tree leaves \cite{Zimm}
    and is studied by lubrication approximation in a model adapted
    to thin layers \cite{gouin7,gouin6}}. It is natural to forecast that the diameters of xylem
    tubes must be the result of a competition between evaporation in  tubes which reduces the flow of sap
    and the flux of transpiration in micropores inducing the motion strength.\newline
Computational methods, such as density-functional theory (DFT) and kinetic Monte Carlo (KMC) have already had major success in nanoscience.
 We have obtained the disjoining pressure and Gibbs energy curves for two different solid walls.
  In cases of  Lifshitz analysis \cite{Lifshitz} and
 van der Waals theory \cite{Henderson}, the disjoining pressure behaviors
  are respectively as $\Pi\sim  h^{-3}$ and $\Pi\sim \exp ( -h)$.
  These two behaviors seem  unable to study a   film with a thickness
   between one and three nanometers.\newline
It is wondering to observe that the density-functional theory
expressed by a   rough model correcting van der Waals' with a
surface density-functional at the walls enables to obtain a good
order of magnitude of the ascent of sap. This result is obtained
without any complex weighted density-functional and without taking
into account the quantum effects  corresponding to less than an
Amgstr\"{o}m length scale.  The surface density-functional at the
wall takes into account
 a power-law behavior associated with a balance between attractive
 and repulsive forces, the square-gradient
 functional schematizing the liquid-vapor interface effects.
 This biophysical observation seems to prove that this
 kind of functional can be a good tool to study models of
 liquids in contact with solids at a small nanoscale range.\\
Due to a remark  by James R. Henderson \cite{HendersonB}, it is interesting to note that \emph{if we switch the micro-tube surfaces to wedge geometry} as in \cite{Finn} or to corrugated surface \emph{then, it is much easier to obtain the complete wetting requirement. Thus, plants can avoid having very high energy
surfaces, but still be internally wet, if they pass liquids through wedge shaped corrugated pores. The wedge does not have to be perfect on the nanometer scale to significantly enhance the amount of liquid that would be passed at modest pressures corresponding to nano-sized planar films. It is bound to improve on the calculation because it enhances the surface to volume ratio}.\\
In such a case, we remark that the wall boundary can always be considered as a plane surface with an average surface energy as in Wenzel's formula \cite{Wenzel}.

\bigskip{\small {\ \textbf{Acknowledgment}: {
The author dedicates the paper to the memory of Professor Pierre
Casal, his Master and friend. }}


\begin{thebibliography}{99}


\bibitem{Flindt} Flindt R.: Amazing Numbers in Biology, Springer, Berlin  (2006).

\bibitem{Koch} Koch W., Sillett S.C., Jennings G.M., Davis S.D.: The
limit to tree height, Nature   \textbf{428},   851-854 (2004).

\bibitem{Zimm} Zimmermann M.H.: Xylem Structure and the Ascent of
Sap, Springer, Berlin  (1983).

\bibitem{Dixon} Dixon H.H., Joly J.: On the ascent of sap, Phil.
Trans. Roy. Soc. London, B \textbf{186},  563-576 (1894).

\bibitem{Honert} van der Honert T.H.: Water transport in plants as a
catenary process, Discussions of the Faraday Society  \textbf{3},
1105-1113 (1948).

\bibitem{Tyree1} Tyree M.T., Sperry J.S.:  The vunerability of xylem
to cavitation and embolism, Annu. Rev. Plant Physio. Plant Mol.
Biol.  \textbf{40}, 19-38 (1989).


\bibitem{Zimm2} Zimmermann U., Schneider H.,  Wegner L.H., Haase A.:
Water ascent in tall trees: does evolution of land plants rely on a
highly metastable state? Tansley review, New Phytologist
\textbf{162}, 575-615 (2004).

\bibitem{Preston} Preston R.D.:  Movement of water in higher plants. In:
Deformation and Flow in Biological Systems, Ed. Frey-Wyssling A,
North Holland Publishing, Amsterdam, 257-321 (1952).


\bibitem{Mackay} Mackay J.F.G., Weatherley P.E.: The effects of transverse cuts
through the stems of transpiring woody plants on water transport and
stress in the leaves, J. of Exp. Botany \textbf{24}, 15-28 (1973).

\bibitem{Eisenhut} Eisenhut G.: Neue Erkenntnisse über den Wassertransport in Bäumen,
Holz–Zentralblatt \textbf{55}, 851-853 (1988).

\bibitem{Benkert} Benkert R, Balling A, Zimmermann U.:  Direct measurements of the
pressure and flow in the xylem vessels of nicotiana tabacum and
their dependence on flow resistance and transpiration rate, Botanica
Acta \textbf{104}, 423-432 (1991).

\bibitem{Balling} Balling A., Zimmermann U.: Comparative
measurements of the xylem pressure of nicotiana plants by means of
the pressure bomb and pressure probe, Planta  \textbf{182}, 325-338
(1990).


\bibitem{Zimmu} Zimmermann U., Haase A., Langbein D., Meinzer F.:
Mechanism of long-distance water transport in plants: a
re-examination of some paradigms in the light of new evidence, Phil.
Trans. Roy. Soc. London  \textbf{431}, 19-31 (1993).

\bibitem{Tyree2} Tyree M.T.: The cohesion-tension theory of sap
ascent: current contreversies, J. Exp. Botany, \textbf{48},
1753-1765 (1997).


\bibitem{Isola} dell'Isola F.,  Gouin H., Rotoli G., Nucleation of spherical
shell-like interfaces by second gradient theory: numerical
simulations,
 Eur. J.  Mech., B/Fluids,
\textbf{15}, 4, pp. 545-568 (1996).

\bibitem{Derjaguin}  {Derjaguin B.V., Chuarev N.V., Muller V.M.}: {Surfaces
Forces}, {Plenum Press, New York (1987).}

\bibitem{Israel}  Israelachvili J.: Intermolecular Forces, Academic Press,
New York  (1992).

\bibitem{Lifshitz} Dzyaloshinsky I.E., Lifshitz E.M., Pitaevsky
L.P.: The general theory of van der Waals forces, Adv. Phys., \textbf{10},
165-209 (1961).


\bibitem{Ono}  {Ono S., Kondo S.,} {Molecular theory of surface
tension in
liquid}. In: Structure of Liquids, S. Fl\"{u}gge (ed.){%
Encyclopedia of Physics, X,} {Springer verlag, Berlin  (1960)}.

\bibitem{gouin7} Gouin H., Gavrilyuk S.:
arXiv:0809.3489,
 Dynamics of liquid
nanofilms, Int. J. Eng. Sci. (2008) doi: 10.1016/j.engsci
2008.05.002 (in press).

\bibitem{Bhushan}  Bhushan B.: Springer Handbook of Nanotechnology,
Springer, Berlin, (2004).


\bibitem{vdW}  {van der Waals J.D.}: {Thermodynamique de la capillarit\'{e}
dans l'hypoth\`{e}se d'une variation continue de densit\'{e},}
{Archives N\'{e}erlandaises} {\textbf{28}}, { {121-209}
(1894-1895)}.


\bibitem{Widom}  Widom B.: What do we know that van der Waals did not know?,
Physica A  \textbf{263}, 500-515 (1999).

\bibitem{Rocard}  Rocard Y.: Thermodynamique, Masson, Paris  (1952).


\bibitem{chernov1} Chernov A.A.,  Mikheev L.V.:  Wetting of solid
surfaces by a structured simple liquid: Effect of fluctuations,
Phys. Rev. Lett. \textbf{60}, 2488 - 2491 (1988).

\bibitem{chernov2} Chernov A.A., Mikheev L.V.: Wetting and surface melting:
Capillary fluctuations versus layerwise short-range order,  Physica
A  \textbf{157},  1042-1058  (1989).

\bibitem{Evans1}  Evans R.: The nature of liquid-vapour interface and other topics in the statistical mechanics of non-uniform classical fluids, Adv. Phys. \textbf{28}, 143-200 (1979).

\bibitem{Evans3} Evans R., Leote de Carvalho J.F.,
Henderson J.R., Hoyle D.C.: Asymptotic decay of correlations in
liquids and their mixtures, J. Chem. Phys.  \textbf{100}, 591-603
(1994)

\bibitem{Fisher} Fisher M.E., Jin A.J.: Effective potentials, constraints,
and critical wetting theory, Phys. Rev. B  \textbf{44}, 3, 1430 -
1433 (1991).

\bibitem{Henderson} Henderson J.R.: Statistical mechanics of the disjoining pressure of a planar
 film, Phys. Rev. E  \textbf{72}, 051602 (2005).

\bibitem{rowlinson}  {Rowlinson J.S., Widom B.:} {Molecular Theory of
Capillarity,} {Clarendon Press, Oxford  (1984)}.

\bibitem{Fisher1} Nakanishi H., Fisher M.E., Multicriticality of wetting, prewetting,
and surface transitions, Phys. Rev. Lett. \textbf{49}, 1565 - 1568 (1982).

\bibitem{Cahn0}  Cahn J.W.: Critical point wetting, J. Chem. Phys. \textbf{66},
3667-3672  (1977).

\bibitem{gouin}  {Gouin H.}: arXiv:0801.4481, Energy of interaction between solid surfaces
and liquids, J. Phys. Chem. B \textbf{102}, 1212-1218 (1998).


\bibitem{gouin4}  Gouin H.: Utilization of the second gradient theory in
continuum mechanics to study the motion and thermodynamics of
liquid-vapor interfaces, Physicochemical Hydrodynamics, B Physics
\textbf{174},  667-682 (1987).

\bibitem{Gouin1}  Gouin H., Kosi\'{n}ski W.:  arXiv:0802.1995, Boundary conditions for a
capillary fluid in contact with a wall, Arch. Mech. \textbf{50},
907-916 (1998).

\bibitem{seppecher} Seppecher P.: Equilibrium of a Cahn and Hilliard fluid on a wall:
influence of the wetting properties of the fluid upon stability of a
thin liquid film, Eur. J. Mech., B/fluids  \textbf{12}, 61-84
(1993).

\bibitem{de Gennes1}  de Gennes P.G.: Wetting : statics and dynamics, Rev.
Mod. Phys. \textbf{57},  827-863 (1985).

\bibitem{de Gennes2}    de Gennes P.G., Brochard-Wyart F., Qu\'{e}r\'{e} D.:
Capillarity and Wetting Phenomena: Drops, Bubbles, Pearls, Waves,
Springer, New York  (2004).

\bibitem{Sheludko} Sheludko A.: Thin liquid films, Adv. Colloid Interface Sci. \textbf{1}, 391-464 (1967).

\bibitem{Lutsko} Lutsko J.M.: Density functional theory of inhomogeneous liquids. I.
 The liquid-vapor interface in Lennard-Jones fluids, J. Chem. Phys. \textbf{127}, 054701 (2007).

\bibitem{Gavrilyuk} Gavrilyuk S.L., Akhatov I.:
 Model of a liquid nanofilm on a solid substrate based on the van der Waals concept of capillarity,
  Phys. Rev. E. \textbf{73}, 021604 (2006).

\bibitem{Pismen} Pismen L.M., Pomeau Y.:
Disjoining potential and spreading of thin liquid layers in the
diffuse-interface model coupled to hydrodynamics, Phys. Rev. E.
\textbf{62}, 2480-2492 (2000).


\bibitem{trusk} {Truskinovsky L.}: Equilibrium phase boundaries,
Sov. Phys. Dokl. \textbf{27}, 551-553 (1982).

\bibitem{espanet}  Gouin H., Espanet L.: arXiv:0807.5023, Bubble number in a caviting flow, {%
Comptes Rendus Acad. Sci. Paris} {\textbf{328}, IIb}  {151-157} (2000).

\bibitem{Handbook}  Handbook of Chemistry and Physics, 65th Edition, CRC
Press, Boca Raton  (1984-1985).


\bibitem{gouin6} Gouin H.: arXiv:0809.2346, A mechanical model for liquid nanolayers,
in: Waves and Stability in Continuous Media, Eds. N. Manganaro, R.
Monaco \& S. Rionero, World Scientific, Singapore, 306-315 (2008).

\bibitem{HendersonB} Henderson J.R.: Private communication and discussion (September 2008).

\bibitem{Finn} Concus P., Finn R.: On the behavior of a capillary surface in a wedge, Proc. Nat. Acad. Sci. \textbf{63}, 292-299 (1969).

\bibitem{Wenzel}  Wenzel  T.N.: Surface roughness and contact angle. J. Phys. Colloid. Chem.
\textbf{53}, 1466
(1949).
\end{thebibliography}
\end{document}